\documentclass[aps,prl,twocolumn,letterpaper,groupedaddress,10pt]{revtex4-2}
\usepackage{amssymb,amsthm,amsmath,amsfonts}
\usepackage{graphicx,ulem,enumerate,bbm,bm}
\usepackage[pdftex,dvipsnames,usenames]{xcolor}
\usepackage[colorlinks=true,urlcolor=blue,citecolor=blue,linkcolor=blue]{hyperref}
\usepackage{subfiles}

\begin{document}

\title{Local-oscillator-agnostic squeezing detection}

\author{Suchitra Krishnaswamy}
    \affiliation{Paderborn University, Institute for Photonic Quantum Systems (PhoQS), Warburger Stra\ss{}e 100, D-33098 Paderborn, Germany}
    \affiliation{Paderborn University, Department of Physics, Warburger Stra\ss{}e 100, D-33098 Paderborn, Germany}

\author{Dhrithi Maria}
    \affiliation{Paderborn University, Institute for Photonic Quantum Systems (PhoQS), Warburger Stra\ss{}e 100, D-33098 Paderborn, Germany}
    \affiliation{Paderborn University, Department of Physics, Warburger Stra\ss{}e 100, D-33098 Paderborn, Germany}

\author{Laura Ares}
	\email{laura.ares.santos@uni-paderborn.de}
    \affiliation{Paderborn University, Institute for Photonic Quantum Systems (PhoQS), Warburger Stra\ss{}e 100, D-33098 Paderborn, Germany}
    \affiliation{Paderborn University, Department of Physics, Warburger Stra\ss{}e 100, D-33098 Paderborn, Germany}
\author{Lorenzo M. Procopio}
    \affiliation{Paderborn University, Institute for Photonic Quantum Systems (PhoQS), Warburger Stra\ss{}e 100, D-33098 Paderborn, Germany}
    \affiliation{Paderborn University, Department of Physics, Warburger Stra\ss{}e 100, D-33098 Paderborn, Germany}

\author{Tim J. Bartley}
    \affiliation{Paderborn University, Institute for Photonic Quantum Systems (PhoQS), Warburger Stra\ss{}e 100, D-33098 Paderborn, Germany}
    \affiliation{Paderborn University, Department of Physics, Warburger Stra\ss{}e 100, D-33098 Paderborn, Germany}
    
\author{Jan Sperling}
    \affiliation{Paderborn University, Institute for Photonic Quantum Systems (PhoQS), Warburger Stra\ss{}e 100, D-33098 Paderborn, Germany}
    \affiliation{Paderborn University, Department of Physics, Warburger Stra\ss{}e 100, D-33098 Paderborn, Germany}

\date{\today}

\begin{abstract}
	We address the problem of measuring nonclassicality in continuous-variable bosonic systems without having access to a known reference signal.
	To this end, we construct broader classes of criteria for nonclassicality that allow us to investigate quantum phenomena regardless of the quantumness of selected subsystems.
	Such witnesses are based on the notion of partial normal ordering.
	This approach is applied to balanced homodyne detection using arbitrary, potentially nonclassical local oscillator states, yet only revealing the probed signal's quantumness.
	Our framework is compared to standard techniques, and the robustness and advanced sensitivity of our approach are shown.
	Therefore, a widely applicable framework, well-suited for applications in quantum metrology and quantum information, is derived to assess the quantum features of a photonic system when a well-defined coherent laser as a reference state is not available in the physical domain under study.
\end{abstract}

\maketitle

\paragraph*{Introduction.---}

	The quantum nature of light remains one of the most important concepts that advances our understanding of fundamental quantum effects and their exploitation in today's quantum technologies \cite{FT20,W21,Petal22}.
	This has contributed to further various fields, such as quantum metrology \cite{B22}, quantum computation \cite{BL05,ARL14}, quantum error correction \cite{GKP01}, gravitational wave detection \cite{SHSD04,Jetal24}, etc.
	Thus, measuring the quantum properties of a signal (SI) light field is paramount in contemporary research.

	Homodyne detection is an ubiquitous technique to fully characterize the quantum nature of light \cite{WVO99,LR09}.
	Specifically, field fluctuations below classical shot noise---dubbed squeezing \cite{W83,WKHW86}---became a seminal hallmark of a phase-sensitive quantum effect that is commonly measured via balanced homodyne detection.
	Some theoretical and experimental schemes consider different photodetection approaches \cite{SVA15,LSV15,PSSB24},
	and some are even photon-detector-agnostic versions of the homodyning measurement protocol that have been recently devised and implemented \cite{Setal20}.
	Still, there is a dependence of this scheme on a coherent state as a phase reference, the so-called local oscillator (LO).
	Highly interesting works have successfully explored the substitution of the coherent state with other, nonclassical Gaussian \cite{K97} and non-Gaussian \cite{CL22} LOs.
	However, this still demands that the LO is known, for example, through a prior characterization.
	By comparison, in entanglement theory, device-independent entanglement witnesses exists, which is a highly sought-after property as it allows one to certify sophisticated quantum correlations without relying on the features of a measurement device \cite{SU01,BGLP11}.

	Thus, the question arises if there is a way to detect squeezing of an unknown SI with an unknown LO.
	For example, in an experimentally interesting regime, there might not be laser sources available from which one can derive a coherent-state LO.
	Note that producing suitable LOs constitutes its own field of multidisciplinary research \cite{HPSKMTBRL05,MWGJSCMEM04,CG96,KK94,BCKCJ67}.
	The main problem with an LO-agnostic approach that one can identify is that measured squeezing might not be a property of the probed SI light, but a feature that originates from the unknown LO;
	this can result in false-positive nonclassicality tests when characterizing the SI.
	Here, we give an affirmative answer about the existence of LO-agnostic schemes by constructing squeezing criteria, which overcome the aforementioned limitations.

	In this contribution, we first establish the family of two-mode states whose nonclassical nature is independent from the second, auxiliary mode, which is the stand-in for the LO.
	We then derive a general methodology to formulate witnesses whose negative expected values are the signature of the first mode's nonclassicality and cannot stem from the auxiliary mode.
	Variance-based criteria as a phase-sensitive nonclassicality witness are obtained as a special case, allowing for the application in a balanced homodyne detection setup with arbitrary classical and nonclassical LOs.
	We compare our approach with the common coherent-LO-based measurement, demonstrate improved noise suppression behavior, and investigate the robustness of our method against typical experimental imperfections.
	For a given SI, the LO optimization for highest noise suppression is discussed, targeting applications in quantum metrology.

\paragraph*{Quantum resource model.---}

	We consider two distinct bosonic (e.g., optical) modes, dubbed $A$ and $B$.
	As a classical pure state, the signal $A$ is considered to be a coherent state \cite{S26,H85}, $|\alpha\rangle\in\mathcal H_A$ for $\alpha\in\mathbb C$, and the state in the auxiliary system $B$ is not constrained, $|\psi\rangle\in\mathcal H_B$.
	The closure of the convex hull of such pure states yields the classical mixed states of the composite system,
	\begin{equation}
		\label{eq:Definition}
		\hat\rho_\mathrm{cl.}=
		\int dP(\alpha,\psi) |\alpha\rangle\langle\alpha|\otimes |\psi\rangle\langle\psi|,
	\end{equation}
	where $P\geq0$ is a probability measure over $\mathbb C\times\mathcal H_B$.
	A state $\hat\rho$ is deemed nonclassical in this model if it does not admit the form in Eq. \eqref{eq:Definition}, i.e., $\hat\rho\neq\hat\rho_\mathrm{cl.}$;
	see also Ref. \cite{ASCBZV17} in this context.
	Specifically, this means that the quantumness of subsystem $B$ does not affect the nonclassicality defined here.
	Rather, a nonclassical state $\hat\rho$ in our definition exhibits nonclassicality in $A$ or includes nonclassical correlations between $A$ and $B$ or both.
	Later, we consider states of the form $\hat\rho_A\otimes\hat\rho_B$, disallowing the second and third options.

\paragraph*{Witnessing subsystem nonclassicality.---}

	The following construction is motivated through the necessary and sufficient family of normally ordered nonclassicality criteria \cite{SRV05,MBWLN10} and the partial transposition criterion \cite{P96,HHH96} from entanglement theory, only altering one subsystem.
	In particular, we take a generic two-mode operator expansion $\hat f=\sum_j f_j \hat A_j \otimes\hat B_j$ and define the partial normal order as
	\begin{equation}
		\label{eq:PartialNormalOrder}
		{\stackrel{A}{:}}\hat f^\dag\hat f{\stackrel{A}{:}}
		=\sum_{j,j'} f_j^\ast f_{j'}
		{:}\hat A_j^\dag\hat A_{j'}{:}\otimes \hat B_j^\dag\hat B_{j'},
	\end{equation}
	where ${:}\cdots{:}$ denotes the common normal-ordering prescription \cite{AW70,VW06} and ${\stackrel{A}{:}}\cdots{\stackrel{A}{:}}$ specifies that normal ordering is carried out only with respect to subsystem $A$, not $B$.
	One can show that the operator in Eq. \eqref{eq:PartialNormalOrder} takes nonnegative values for our classical states in Eq. \eqref{eq:Definition},
	\begin{equation}
		\label{eq:ExpectedPartNO}
	\begin{aligned}
		{}&
		\left\langle{\stackrel{A}{:}}\hat f^\dag\hat f{\stackrel{A}{:}}\right\rangle
		=
		\sum_{j,j'} f_j^\ast f_{j'}
		\mathrm{tr}\left(
			\hat\rho_\mathrm{cl.}
			{:}\hat A_j^\dag\hat A_{j'}{:}\otimes \hat B_j^\dag\hat 1\hat B_{j'}
		\right)
		\\
		={}&
		\int dP(\alpha,\psi)\sum_{n=0}^\infty
		\left|
			\sum_{j'} f_{j'}A_{j'}(\alpha) \langle n|\hat B_{j'}|\psi\rangle
		\right|^2
		,
	\end{aligned}
	\end{equation}
	using the identity $\hat 1=\sum_{n=0}^\infty |n\rangle\langle n|$ and the abbreviation $\langle\alpha|{:}\hat A_j{:}|\alpha\rangle=A_j(\alpha)$, also yielding $\langle\alpha|{:}\hat A_{j}^\dag\hat A_{j'}{:}|\alpha\rangle=A_{j}(\alpha)^\ast A_{j'}(\alpha)$ \cite{AW70,VW06}.
	Because of $P\geq0$, the expression in Eq. \eqref{eq:ExpectedPartNO} is nonnegative, implying the classical constraint
	\begin{equation}
		\label{eq:ClassicalConstraint}
		0
		\stackrel{\text{cl.}}{\leq}
		\left\langle{\stackrel{A}{:}}\hat f^\dag\hat f{\stackrel{A}{:}}\right\rangle.
	\end{equation}
	Conversely, $0>\langle{\stackrel{A}{:}}\hat f^\dag\hat f{\stackrel{A}{:}}\rangle$ certifies that the probed state is nonclassical regardless of the states in $B$.
	As an essential example, let $\hat f=\hat L-\langle{\stackrel{A}{:}}\hat L{\stackrel{A}{:}}\rangle \hat 1\otimes\hat 1$ for an arbitrary, self-adjoint observable $\hat L$.
	Expanding the criterion in Eq. \eqref{eq:ClassicalConstraint} then reads as a partially normal-ordered variance,
	\begin{equation}
		\label{eq:VarianceConstraint}
		0
		\stackrel{\text{cl.}}{\leq}
		\left\langle{\stackrel{A}{:}}\hat L^2{\stackrel{A}{:}}\right\rangle
		-\left\langle{\stackrel{A}{:}}\hat L{\stackrel{A}{:}}\right\rangle^2
		=
		\left\langle{\stackrel{A}{:}}(\Delta\hat L)^2{\stackrel{A}{:}}\right\rangle
		.
	\end{equation}

\paragraph*{LO-agnostic squeezing criterion.---}

\begin{figure}
	\includegraphics[width=.9\columnwidth]{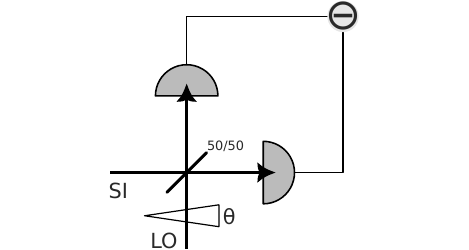}
	\caption{%
		Balanced homodyne detection.
		The SI and the here arbitrary LO interfere on a $50/50$ beam splitter.
		Both outputs are measured with photodetectors, whose difference defines the observable $\hat L$.
		The LO's phase $\theta$ may be controlled.
	}\label{fig:BHD}
\end{figure}

	In Fig. \ref{fig:BHD}, the measurement setting under study is described.
	In contrast to standard balanced homodyne detection \cite{WVO99,LR09}, we do not assume that we have access to a coherent states---or any specific state for that matter---as our LO in the auxiliary system $B$.
	Still, the goal is witness the presence of squeezing in the SI, in mode $A$.
	The measured photon-number difference, after superimposing the SI and LO on a $50/50$ beam splitter and allowing for phase ($\theta$) control of the LO, can be recast into the observable
	\begin{align}
		\nonumber
		\hat L
		={}&
		\left(\frac{\hat a{+}e^{i\theta}\hat b}{\sqrt 2}\right)^\dag
		\left(\frac{\hat a{+}e^{i\theta}\hat b}{\sqrt 2}\right)
		-
		\left(\frac{\hat a{-}e^{i\theta}\hat b}{\sqrt 2}\right)^\dag
		\left(\frac{\hat a{-}e^{i\theta}\hat b}{\sqrt 2}\right)
		\\
		\label{eq:LinInterference}
		={}&
		e^{i\theta}\hat a^\dag\hat b
		+e^{-i\theta}\hat a\hat b^\dag,
	\end{align}
	using the annihilation operators $\hat a$ and $\hat b$ for the SI mode $A$ and LO mode $B$, respectively.
	Note that $\hat L$ is in normal order;
	i.e., $\hat L={:}\hat L{:}={\stackrel{A}{:}}\hat L{\stackrel{A}{:}}$ holds true.

	For assessing the fluctuations in Eq. \eqref{eq:VarianceConstraint}, one computes
	\begin{equation}
		\hat L^2
		=
		e^{2i\theta}\hat a^{\dag 2}\hat b^2
		+e^{-2i\theta}\hat a^\dag\hat b^{\dag 2}
		+\hat a^\dag\hat a\hat b\hat b^\dag
		+\hat a\hat a^\dag\hat b^\dag\hat b,
	\end{equation}
	which is neither in partial nor in full normal order.
	But the full and partial ordering of this can be determined straightforwardly, resulting in fluctuations of the form
	\begin{subequations}
	\begin{align}
		\label{eq:PartNOVariance}
		\left\langle{\stackrel{A}{:}}
			(\Delta\hat L)^2
		{\stackrel{A}{:}}\right\rangle
		={}&
		\left\langle
			(\Delta\hat L)^2
		\right\rangle
		-\langle \hat b^\dag\hat b\rangle
		\quad\text{and}
		\\
        \label{eq:PartNOVariance_b}
		\left\langle{:}
			(\Delta\hat L)^2
		{:}\right\rangle
		={}&
		\left\langle
			(\Delta\hat L)^2
		\right\rangle
		-\langle \hat a^\dag\hat a\rangle
		-\langle \hat b^\dag\hat b\rangle.
	\end{align}
	\end{subequations}
	See the supplemental material \cite{SuppMat} for technical details.
	Note that the common nonclassical criterion that depends on the classicality of $B$ reads $0>\langle{:}
			(\Delta\hat L)^2
		{:}\rangle$
	\cite{SRV05}, which is distinct from violating Eq. \eqref{eq:VarianceConstraint}.
	Except for the assumed existence of moments up to the second order, no assumptions have been made about LO state, including its Gaussian \cite{K97} or non-Gaussian \cite{CL22} nature.

\paragraph*{Shot-noise quantification.---}

	Blocking the SI in Fig. \ref{fig:BHD}, we measure vacuum, directly resulting in
	$\langle
		(\Delta\hat L)^2
	\rangle_\mathrm{vac.}
	=
	\langle \hat b^\dag\hat b\rangle$ \cite{SuppMat}.
	This renders it possible to determine the noise suppression in decibel by manipulating Eq. \eqref{eq:VarianceConstraint}, together with Eq. \eqref{eq:PartNOVariance},
	\begin{equation}
		\label{eq:NoiseParameter}
		\langle b^\dag\hat b\rangle
		\stackrel{\text{cl.}}{\leq}
		\langle(\Delta\hat L)^2\rangle
		\text{ $\Leftrightarrow$ }
		0\stackrel{\text{cl.}}{\leq}
		10\log_{10}\frac{
			\langle
				(\Delta\hat L)^2
			\rangle
		}{
			\langle
				(\Delta\hat L)^2
			\rangle_\mathrm{vac.}
		}=\mathcal N.
	\end{equation}
	Interestingly, this is the same approach as assessing squeezing in common experiments with a coherent LO, and no adjustment of that procedure has to be made with our updated approach.
	The quantity $\mathcal N$ is a noise parameter, which is zero for the shot noise of vacuum.
	$\mathcal N>0$ implies excess noise, and $\mathcal N<0$ implies nonclassicality, i.e., a suppression below vacuum shot noise.

\paragraph*{Remark.---}

	Because of its generality, the inequality in Eq. \eqref{eq:VarianceConstraint} holds true even when interference between the SI and LO is achieved by means other than a linear $50/50$ beam splitter, such as used in unbalanced homodyning \cite{WV96,TJS17} and nonlinear interferometry \cite{GSAJL18,MFTC20}.
	In such a case, the operator $\hat L$ in Eq. \eqref{eq:LinInterference} can be replaced with another, more general functional $F$, i.e., $\hat L=F(\hat a,\hat a^\dag,\hat b,\hat b^\dag)$.
	Although not further discussed here, this also shows that the method put forward applies regardless the choice of the optical interferometer used in experiments.
    For the same reason, an extension to detect higher-order squeezing \cite{HM85} can be developed by properly modifying the general function defining $\hat{L}$ \cite{PY12}.

\paragraph*{Application, robustness, and comparison.---}

	For our applications, we now additionally assume that the LO and the SI are uncorrelated, i.e., $\hat\rho=\hat\rho_\mathrm{SI}\otimes\hat\rho_\mathrm{LO}$.
	The above expressions solely depend on first-order and second-order field correlations, which now can be recast into the form \cite{SuppMat}
	\begin{equation}
		\label{eq:CovarForm}
	\begin{aligned}
		\left\langle
			(\Delta\hat L)^2
		\right\rangle
		={}&
		\mathrm{tr}\left(
			C R_\theta^\mathrm{T} C' R_\theta
		\right)-\frac{1}{2}
		\\
		{}&
		+\xi^\mathrm{T} R_\theta^\mathrm{T} C' R_\theta \xi
		+\xi^{\prime\mathrm{T}} R_\theta C R_\theta^\mathrm{T} \xi^\prime,
	\end{aligned}
	\end{equation}
	where $R_\theta$ is a phase-space rotation matrix, $C$ and $C'$ are the covariance matrices of the SI and LO, respectively, and $\xi$ and $\xi^\prime$ are vectors that include the $x$ and $p$ displacements of the SI and LO.
	In addition, we have
	\begin{subequations}
	\begin{align}
		\langle \hat a^\dag\hat a\rangle
		={}&
		\frac{
			\mathrm{tr}(C)
			-1
			+\xi^\mathrm{T}\xi
		}{2}
		\quad\text{and}
		\\
		\langle \hat b^\dag\hat b\rangle
		={}&
		\frac{
			\mathrm{tr}(C')
			-1
			+\xi^{\prime\mathrm{T}}\xi^\prime
		}{2}.
	\end{align}
	\end{subequations}
	Although we are using squeezed states in the following examples, we again emphasize that the (co-)variances used can come from arbitrary states.

\begin{figure}
	\includegraphics[width=.9\columnwidth]{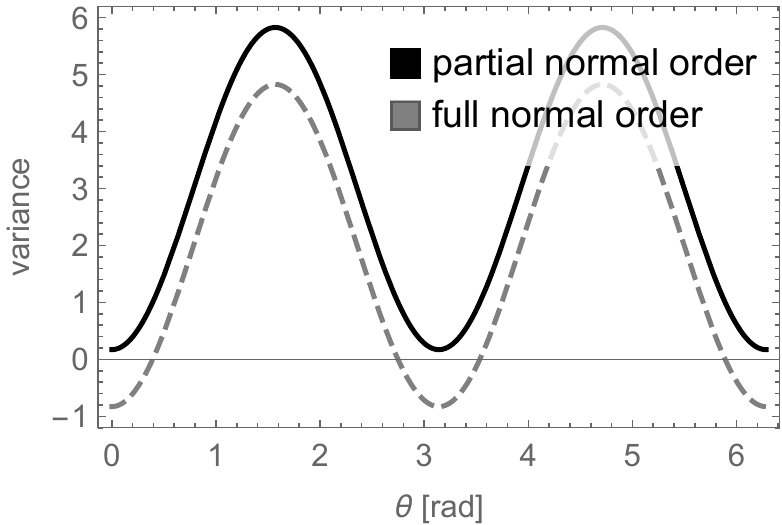}
	\caption{%
		Partially normally ordered $\left\langle{\stackrel{A}{:}}(\Delta\hat L)^2{\stackrel{A}{:}}\right\rangle$ (solid, black) and fully normally ordered $\left\langle:
			(\Delta\hat L)^2
		:\right\rangle$ (dashed, gray) field fluctuations as a function of the phase parameter $\theta$.
		Here, the LO is a  squeezed-vacuum state ($3\,\mathrm{dB}$), and a classical coherent state (mean photon number of one) is taken as the SI.
		The full normal ordering becomes negative not because of a nonclassical SI but because of the nonclassical LO, yielding a false-positive squeezing indicator for the SI when the LO's properties are not known.
		Our method based on partial normal ordering, however, correctly shows that no nonclassicality is in the SI.
	}\label{fig:Fluctuations}
\end{figure}

	Figure \ref{fig:Fluctuations} shows the application of our approach in comparison with the standard technique, using a reversed scenario of a nonclassical LO and a classical SI.
	The common criterion for nonclassicality, $0>\langle{:}(\Delta\hat L)^2{:}\rangle=\langle(\Delta\hat L)^2\rangle-\langle\hat a^\dagger\hat a\rangle-\langle\hat b^\dagger\hat b\rangle$, shows fake negativities despite a classical SI as it cannot separate SI and LO quantum properties.
	By contrast, our LO-agnostic approach complies with the classical constraint in Eq. \eqref{eq:VarianceConstraint}, $0\leq\langle{\stackrel{A}{:}}(\Delta\hat L)^2{\stackrel{A}{:}}\rangle=\langle(\Delta\hat L)^2\rangle-\langle\hat b^\dag\hat b\rangle$, correctly identifying the classical behavior of the SI state under study.

\begin{figure}
	\includegraphics[width=.9\columnwidth]{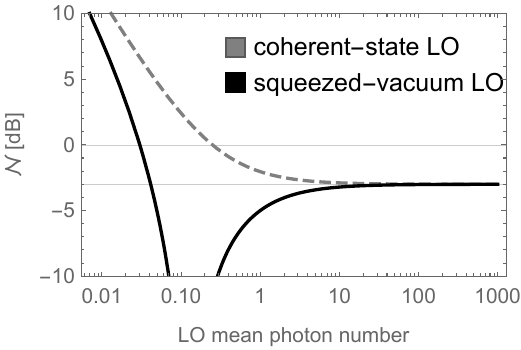}
	\caption{%
		Noise parameter $\mathcal N$ [Eq. \eqref{eq:NoiseParameter}] as a function of the LO intensity $\langle\hat b^\dag\hat b\rangle$ (logarithmic scale) for the common coherent-state LO (dashed, gray) and a squeezed-vacuum LO (solid, black).
		The nonclassical SI ($3\,\mathrm{dB}$ squeezing) yields a negative value $\mathcal N<0$, thus demonstrates noise suppression below the shot-noise limit (at $\mathcal N=0$).
		A coherent LO laser underperforms, requiring a high intensity and approaching the highest sensitivity (also $3\,\mathrm{dB}$) from above.
		A squeezed LO outperforms the coherent one, including an optimal noise reduction at a finite squeezing value (again $3\,\mathrm{dB} \rightarrow \langle\hat b^\dag \hat b\rangle\approx0.124$).
	}\label{fig:NoiseLevel}
\end{figure}

	In Fig. \ref{fig:NoiseLevel}, we study the noise suppression $\mathcal N<0$ [Eq. \eqref{eq:NoiseParameter}] to determine the sensitivity of our nonclassicality test for a squeezed SI ($3\,\mathrm{dB}$) with our LO-agnostic approach, using a coherent LO (dashed curve, $|\beta\rangle$) and a squeezed-vacuum LO (solid curve, $|\zeta'\rangle$).
	The noise level $\mathcal N$ is plotted as a function of the mean photon number of the LO, being $\langle\hat b^\dag\hat b\rangle=|\beta|^2$ and $\langle\hat b^\dag\hat b\rangle=\sinh^2\zeta'$ for the two LOs, over several orders of magnitude.
	A strong coherent LO is needed to approach to maximnal noise reduction of $\langle (\Delta\hat L)^2\rangle$, which is at most the SI's squeezing level itself, $\mathcal N\to-3$ for $|\beta|^2\to\infty$.
	By contrast, a moderately squeezed LO yields a perfect noise reduction, $\langle (\Delta\hat L)^2\rangle=0$ and $\mathcal N=-\infty$.
	This can be deduced from Eq. \eqref{eq:CovarForm} as follows:
	Suppose a diagonal and minimal-uncertainty covariance matrix $C=\mathrm{diag}(e^{-2\zeta}/2,e^{2\zeta}/2)$ as the squeezed SI $|\zeta\rangle$ and $C'=\mathrm{diag}(e^{-2\zeta'}/2,e^{2\zeta'}/2)$ for the squeezed LO $|\zeta'\rangle$---together with zero displacements $\xi$ and $\xi'$ for both.
	Let $\zeta'=\zeta$ and $\theta=90^\circ$ (likewise $\zeta'=-\zeta$ and $\theta=0$), which yields $\langle (\Delta\hat L)^2\rangle=e^{-2(\zeta-\zeta)}/4+e^{2(\zeta-\zeta)}/4-1/2=0$ in Eq. \eqref{eq:CovarForm}, therefore $\mathcal N=10\log_{10}(0)=-\infty$ in Eq. \eqref{eq:NoiseParameter}.
	Such a (here perfect) noise supression below the classical shot-noise limit with a limited energy demand is not attainable with coherent-state LOs and is highly useful for applications in quantum metrology.

	The above examples assumed ideal systems to demonstrate the fundamental function of our approach.
	Now, we discuss the influence of loss and noise on the SI and their impact on the performance of our criteria.
	(See the supplemental material \cite{SuppMat} for the exact calculations.)
	First, if a loss level $1-\eta$ (with quantum efficiency $0\leq\eta\leq 1$) is imposed on the SI state, we then find
	\begin{equation}
		\left\langle{\stackrel{A}{:}}
			(\Delta\hat L)^2
		{\stackrel{A}{:}}\right\rangle
		=
		\eta\left.
		\left\langle{\stackrel{A}{:}}
			(\Delta\hat L)^2
		{\stackrel{A}{:}}\right\rangle
		\right|_{\eta=1},
	\end{equation}
	compared to the ideal case $\eta=1$.
	Since $\eta>0$ when excluding the trivial case $\eta=0$, no sign change occurs, only a reduction of the overall classical or nonclassical signature occurs.
	Second, we can consider a noise channel, commonly modeled via a bath-coupled amplifier with a gain $g\geq1$;
	this adds excess noise by amplifying and transfering vacuum fluctuations from a coupled bath to the SI.
	We obtain
	\begin{equation}
	\begin{aligned}
		\left\langle{\stackrel{A}{:}}
			(\Delta\hat L)^2
		{\stackrel{A}{:}}\right\rangle
		={}&
		g\left.
		\left\langle{\stackrel{A}{:}}
			(\Delta\hat L)^2
		{\stackrel{A}{:}}\right\rangle
		\right|_{g=1}
		\\
		{}&
		+(g-1)
		\left(
			\langle \hat b\hat b^\dag \rangle
			+ \langle \hat b^\dag\hat b \rangle
		\right).
	\end{aligned}
	\end{equation}
	Similar to loss, a positive scaling without a sign change can be observed for the partially normal-ordered variance.
	However, additive excess noise proportional to $g-1$ and the LO's intensity might hide negativities that would be present without the excess-noise addition, where $g=1$.
	However, it is important to note that negativities cannot be generated through loss and noise, further fostering the robustness against false-positive nonclassicality indicators of our methodology.

\paragraph*{Summary.---}

	We derived an LO-agnostic nonclassicality condition based on first-order and second-order field moments to probe the squeezed nature of quantum light.
	To this end, a resource-theoretic framework was constructed, leading to a much broader class of subsystem nonclassicality witnesses.
	Within this framework, we can determine the quantum properties of a system when coupled to an unknown auxiliary system.
	Then, as a special case, we resolved the open problem of certifying squeezing without the need to characterize the LO in a homodyne detection scenario.
	Thereby, measured nonclassicality cannot be caused by any quantum properties of the LO, removing the possibility of false positives when the LO is nonclassical itself.

	While we focus on two modes, an extension to multimode systems and an arbitrary number of auxiliary systems is straightforward, using partial normal order on the probed system modes only.
	In addition, we can use the same approach for other homodyning scenarios, such as eight-port homodyning \cite{LP96,FF99}, weak-field homodyning \cite{DBJVDBW14,Tetal20}, etc.
	Further, we can combine this LO-agnostic feature with our previously introduced detector-agnostic approaches \cite{Setal17,Setal20}.
	This leads to future research, such as more general state-reconstructions principles beyond witnessing.
	An experiment to implement the full LO-agnostic and photodetector-agnostic theory is currently being developed, based on the setup in Ref. \cite{PSSB24}.

\paragraph*{Acknowledgments.---}
	The authors are grateful to Christine Silberhorn for valuable discussions.
    Partially funded by the European Union (ERC, QuESADILLA, 101042399). The views and opinions expressed are, however, those of the author(s) only and do not necessarily reflect those of the European Union or the European Research Council Executive Agency. Neither the European Union nor the granting authority can be held responsible for them.
    This work was partially supported through the QuantERA project QuCABOoSE.


\end{document}


\title{Supplemental Material
\\
Local-oscillator-agnostic squeezing detection}

\author{Suchitra Krishnaswamy}
    \affiliation{Paderborn University, Institute for Photonic Quantum Systems (PhoQS), Warburger Stra\ss{}e 100, D-33098 Paderborn, Germany}
    \affiliation{Paderborn University, Department of Physics, Warburger Stra\ss{}e 100, D-33098 Paderborn, Germany}

\author{Dhrithi Maria}
    \affiliation{Paderborn University, Institute for Photonic Quantum Systems (PhoQS), Warburger Stra\ss{}e 100, D-33098 Paderborn, Germany}
    \affiliation{Paderborn University, Department of Physics, Warburger Stra\ss{}e 100, D-33098 Paderborn, Germany}

\author{Laura Ares}
    \affiliation{Paderborn University, Institute for Photonic Quantum Systems (PhoQS), Warburger Stra\ss{}e 100, D-33098 Paderborn, Germany}
    \affiliation{Paderborn University, Department of Physics, Warburger Stra\ss{}e 100, D-33098 Paderborn, Germany}

\author{Lorenzo M. Procopio}
    \affiliation{Paderborn University, Institute for Photonic Quantum Systems (PhoQS), Warburger Stra\ss{}e 100, D-33098 Paderborn, Germany}
    \affiliation{Paderborn University, Department of Physics, Warburger Stra\ss{}e 100, D-33098 Paderborn, Germany}

\author{Tim J. Bartley}
    \affiliation{Paderborn University, Institute for Photonic Quantum Systems (PhoQS), Warburger Stra\ss{}e 100, D-33098 Paderborn, Germany}
    \affiliation{Paderborn University, Department of Physics, Warburger Stra\ss{}e 100, D-33098 Paderborn, Germany}

\author{Jan Sperling}
    \affiliation{Paderborn University, Institute for Photonic Quantum Systems (PhoQS), Warburger Stra\ss{}e 100, D-33098 Paderborn, Germany}
    \affiliation{Paderborn University, Department of Physics, Warburger Stra\ss{}e 100, D-33098 Paderborn, Germany}

\date{\today}

\begin{abstract}
	In this supplemental material, we provide technical details and exact calculations that support the findings in the main text.
	In Sec. \ref{sec:Gaussian}, the covariance matrix formalism in the context of our approach is discussed, including its application to `-oscillator-agnostic squeezing criteria.
	In Sec. \ref{sec:Imperfections}, we explicitly consider the experimentally most relevant sources of imperfections, loss and noise.
	We use the common acronyms LO and SI for local oscillator and signal, respectively.
\end{abstract}

\maketitle
\appendix

\section{Covariance matrix formalism}
\label{sec:Gaussian}

	In this section, we present the covariance matrix formalism in quantum optics as it pertains to the problem under study.
	See, e.g., Refs. \cite{VW06,ARL14,FT20} for general and more detailed overviews.
	Specifically, the relation between quadrature variables (i.e., $\hat x$ and $\hat p$) and bosonic field operators (i.e., $\hat a$ and $\hat a^\dag$) is outlined here.
	Further, we then relate the concepts to the observable $\hat L$ for balanced homodyne detection from the main part.

\subsection{Fundamentals}

	We can transform between field and momentum operators and creation and annihilation operators via a unitary map $T$,
	\begin{equation}
		\begin{bmatrix}
			\hat a \\ \hat a^\dag
		\end{bmatrix}
		=
		\underbrace{
			\frac{1}{\sqrt2}
			\begin{bmatrix}
				1 & i \\ 1 & -i
			\end{bmatrix}
		}_{
			=T
		}
		\begin{bmatrix}
			\hat x \\ \hat p
		\end{bmatrix}
		,
	\end{equation}
	also relating the fundamental commutation relations $[\hat a,\hat a^\dag]=1$ and $[\hat x,\hat p]=i$.
	Defining $\Delta\hat q=\hat q-\langle \hat q\rangle$ for an arbitrary operator $\hat q$, the covariance matrix $C$ of a generic quantum state is given through the positive semidefinite matrix
	\begin{equation}
	\begin{aligned}
		\left\langle
		\begin{bmatrix}
			\Delta\hat x \\ \Delta\hat p
		\end{bmatrix}
		\begin{bmatrix}
			\Delta\hat x \\ \Delta\hat p
		\end{bmatrix}^\dag
		\right\rangle
		={}&
		\overbrace{
			\begin{bmatrix}
				\langle (\Delta\hat x)^2\rangle
				&
				\langle \frac{1}{2}\{\Delta\hat x, \Delta\hat p\}\rangle
				\\
				\langle \frac{1}{2}\{\Delta\hat x, \Delta\hat p\}\rangle
				&
				\langle (\Delta\hat p)^2\rangle
			\end{bmatrix}
		}^{=C}
		\\
		{}&+\frac{i}{2}
		\underbrace{
			\begin{bmatrix}
				0 & 1 \\ -1 & 0
			\end{bmatrix}
		}_{=\Omega}
	\end{aligned}
	\end{equation}
	and the symplectic matrix $\Omega$, using the anti-commutator $\{\hat x,\hat p\}=\hat x\hat p+\hat p\hat x$ for rendering $C$ symmetric.
	Analogously, we can define
	\begin{equation}
		A
		=
		\left\langle
		\begin{bmatrix}
			\Delta \hat a\Delta \hat a^\dag
			&
			\Delta \hat a\Delta \hat a
			\\
			\Delta \hat a^\dag\Delta \hat a^\dag
			&
			\Delta \hat a^\dag\Delta \hat a
		\end{bmatrix}
		\right\rangle
		=T\left(
			C+\frac{i}{2}\Omega
		\right) T^\dag,
	\end{equation}
	which is directly related to the covariance matrix.
	Note that the commutation relation $\hat a\hat a^\dag-\hat a^\dag\hat a=1$ can be used to exchange the diagonal elements with different operator orderings as needed;
	a useful relation in this direction is
	\begin{equation}
		T\left(
			C-\frac{i}{2}\Omega
		\right) T^\dag
		=
		\begin{bmatrix}
			\langle \Delta\hat a^\dag \Delta\hat a \rangle
			&
			\langle (\Delta \hat a)^2\rangle
			\\
			\langle (\Delta \hat a^\dag)^2\rangle
			&
			\langle \Delta\hat a \Delta\hat a^\dag \rangle
		\end{bmatrix}.
	\end{equation}
	We can further relate phase transformations $P_\theta$ to phase-space rotations $R_\theta$ as follows:
	\begin{equation}
	\begin{aligned}
		P_\theta=T R_\theta T^\dag,
		{}&\text{ where}
		\\
		P_\theta=
		\begin{bmatrix}
			e^{-i\theta} & 0
			\\
			0 & e^{i\theta}
		\end{bmatrix}
		{}&\text{ and }
		R_\theta
		=
		\begin{bmatrix}
			\cos\theta & \sin\theta
			\\
			-\sin\theta & \cos\theta
		\end{bmatrix}.
	\end{aligned}
	\end{equation}

	A pertinent example is the diagonalization of the covariance matrix
	\begin{equation}
		C=R_\phi^\mathrm{T} D R_\phi,
		\text{ with }
		D=\mathrm{diag}[\sigma_x^2,\sigma_p^2],
	\end{equation}
	through a rotation with an angle $\phi$ and the variances $\sigma_x^2$ and $\sigma_p^2$, where $C+\frac{i}{2}\Omega\geq0$ implies the uncertainty relation $\sigma_x^2\sigma_p^2\geq 1/4$.
	From the above relations, we then find the components of the matrix $A$ as
	\begin{equation}
	\begin{aligned}
		\langle \Delta\hat a\Delta a^\dag\rangle
		={}&
		\frac{\sigma_x^2+\sigma_p^2+1}{2},
		\\
		\langle \Delta\hat a^\dag\Delta a\rangle
		={}&
		\frac{\sigma_x^2+\sigma_p^2-1}{2},
		\\
		\text{and }
		\langle (\Delta \hat a)^2\rangle
		={}&
		e^{2i\theta}\frac{\sigma_x^2-\sigma_p^2}{2}
		=\langle (\Delta \hat a^\dag)^2\rangle^\ast.
	\end{aligned}
	\end{equation}
	We remark that, for a squeezing parameter $\zeta\in\mathbb R$ and a thermal photon number $\bar n\geq0$ background (excess noise), the uncertainty relation is obeyed when setting
	\begin{equation}
		\sigma_x^2=\frac{e^{-2\zeta}}{2}+\bar n
		\text{ and }
		\sigma_p^2=\frac{e^{2\zeta}}{2}+\bar n.
	\end{equation}
	In addition, the displacement vector
	$\xi=\left[\begin{smallmatrix}
		\xi_x \\ \xi_p
	\end{smallmatrix}\right]
	=
	\left\langle
		\left[\begin{smallmatrix}
		\hat x \\ \hat p
	\end{smallmatrix}\right]
	\right\rangle$
	relates to the complex-valued coherent amplitude as $\alpha=\langle \hat a\rangle=(\xi_x+i\xi_p)/\sqrt2$.

\subsection{LO-agnostic squeezing}

	In the main part, we found the observable
	\begin{equation}
		\hat L=e^{i\theta}\hat a^\dag\hat b+e^{-i\theta}\hat a\hat b^\dag.
	\end{equation}
	Together with the number operators $\hat b^\dag \hat b$ and $\hat a^\dag \hat a$, this enables us to investigate nonclassicality of the SI alone and joint SI-LO nonclassicality.
	We assumed that the joint state factorizes as $\hat\rho=\hat\rho_\mathrm{SI}\otimes\hat \rho_\mathrm{LO}$ for the sake of simplicity.

	Some algebra, together with the identities formulated above, allows us to write
	\begin{equation}
	\begin{aligned}
		\langle (\Delta\hat L)^2\rangle
		={}&
		e^{2i\theta}\langle \hat a^{\dag 2}\rangle\langle \hat b^2\rangle
		+e^{-2i\theta}\langle \hat a^2\rangle\langle \hat b^{\dag 2}\rangle
		\\
		{}&
		+\langle\hat a^\dag\hat a\rangle\langle\hat b\hat b^\dag\rangle
		+\langle\hat a\hat a^\dag\rangle\langle\hat b^\dag\hat b\rangle
		\\
		{}&
		-\left(
			e^{i\theta}\langle \hat a^\dag\rangle\langle \hat b\rangle
			+e^{-i\theta}\langle \hat a\rangle\langle \hat b^\dag\rangle
		\right)^2
		\\
		={}&
		\mathrm{tr}\left(
			T\left(C{+}\frac{i}{2}\Omega{+}\xi\xi^\mathrm{T}\right)T^\dag
			T R_\theta^\mathrm{T} T^\dag
		\right.
		\\
		{}&
		\phantom{\mathrm{tr}}\left.
			\times
			T\left(C^\prime{-}\frac{i}{2}\Omega{+}\xi^\prime\xi^{\prime\mathrm{T}}\right)T^\dag
			T R_\theta T^\dag
		\right)
		\\
		{}&
		-\left(
			\xi^\mathrm{T} T^\dag
			T R_\theta^\mathrm{T} T^\dag
			T \xi'
		\right)^2
		\\
		={}&
		\mathrm{tr}\left(
			C R_\theta^\mathrm{T} C' R_\theta
		\right)-\frac{1}{2}
		\\
		{}&
		+\xi^\mathrm{T} R_\theta^\mathrm{T} C' R_\theta \xi
		+\xi^{\prime\mathrm{T}} R_\theta C R_\theta^\mathrm{T} \xi^\prime,
	\end{aligned}
	\end{equation}
	where $C'$ and $\xi'$ respectively denote the phase-space covariance and displacement of the LO.
	Moreover, the expected intensities of the SI and LO fields read
	\begin{equation}
	\begin{aligned}
		\langle \hat a^\dag\hat a\rangle
		={}&
		\frac{
			\mathrm{tr}(C)
			-1 +\xi^\mathrm{T}\xi
		}{2},
		\\
		\langle \hat b^\dag\hat b\rangle
		={}&
		\frac{
			\mathrm{tr}(C')
			-1 +\xi^{\prime\mathrm{T}}\xi^\prime
		}{2}.
	\end{aligned}
	\end{equation}
	For example, a displaced squeezed state with thermal background yields $\langle\hat a^\dag\hat a\rangle=\sinh^2\zeta+\bar n+|\alpha|^2$.
	Combining the previous expressions, we can directly evaluate the classical constraints through
	\begin{equation}
	\begin{aligned}
		\left\langle {\stackrel{A}{:}}
			(\Delta\hat L)^2
		{\stackrel{A}{:}} \right\rangle
		={}&
		\langle (\Delta\hat L)^2\rangle
		-\langle \hat b^\dag\hat b\rangle,
		\\
		\left\langle {:}
			(\Delta\hat L)^2
		{:} \right\rangle
		={}&
		\langle (\Delta\hat L)^2\rangle
		-\langle \hat a^\dag\hat a\rangle
		-\langle \hat b^\dag\hat b\rangle,
	\end{aligned}
	\end{equation}
	which follows from
	${\stackrel{A}{:}}
		\hat a\hat a^\dag\hat b^\dag\hat b
	{\stackrel{A}{:}}
	=
	(\hat a\hat a^\dag-\hat 1)\hat b^\dag\hat b
	$
	and
	${:}
		\hat a^\dag\hat a\hat b\hat b^\dag
		+\hat a\hat a^\dag\hat b\hat b^\dag
	{:}
	=
	\hat a^\dag\hat a(\hat b\hat b^\dag-\hat 1)
	+(\hat a\hat a^\dag-\hat 1)\hat b\hat b^\dag
	$ as simple applications of partial and full normal orderings.

\subsection{Discussion}

	Using the exact identities formulated above, all examples and plots from the main part can be generated straightforwardly.
	For example, when the signal is the vacuum state, we have $\sigma_x^2-\sigma_p^2=1/2$ and $\xi_x=\xi_p=0$, resulting in
	\begin{equation}
		\langle(\Delta\hat L)^2\rangle_\mathrm{vac.}
		=
		\langle \hat b^\dag\hat b\rangle.
	\end{equation}
	This is used to determine the LO-agnostic squeezing parameter in decibel as used in the main text.
	\\

\section{Imperfections}
\label{sec:Imperfections}

	In this section, the impact of common imperfections, attenuation and excess noise, is discussed.
	Both imperfections are modeled as is common, i.e., through coupling to a bath mode.
	Since our approach is LO-agnostic, we only compute the imperfections for the signal.

\subsection{Losses}

	In addition, we recall the well-known transformation via a loss channel, determined through the efficiency $\eta$ ($0\leq \eta \leq 1$), as
	\begin{equation}
	\begin{aligned}
		\langle \hat a\rangle\mapsto \sqrt\eta\langle \hat a\rangle,
		\text{ }
		\langle \hat a^2\rangle\mapsto \eta\langle \hat a\rangle,
		\text{ }
		\langle \hat a^\dag\hat a\rangle\mapsto \eta\langle \hat a^\dag\hat a\rangle,
		\\
		\text{and }
		\langle \hat a\hat a^\dag\rangle
		\mapsto
		\eta
		\langle \hat a\hat a^\dag\rangle+1-\eta.
	\end{aligned}
	\end{equation}
	This is a result from a linear coupling to a bath mode $\hat c$ with a bath being in vacuum, where $\hat a\mapsto \sqrt\eta \hat a+\sqrt{1-\eta}\hat c$.
	Focusing on loss for the SI mode, we thus have
	\begin{equation}
	\begin{aligned}
		\langle (\Delta \hat L)^2 \rangle
		\mapsto{}&
		\eta\langle (\Delta \hat L)^2 \rangle
		+(1-\eta)\langle \hat b^\dag\hat b\rangle
		\\
		\text{$\Rightarrow$ }
		\langle (\Delta \hat L)^2 \rangle-\langle \hat b^\dag\hat b\rangle
		\mapsto{}&
		\eta\left(
			\langle (\Delta \hat L)^2 \rangle - \langle\hat b^\dag\hat b\rangle
		\right),
	\end{aligned}
	\end{equation}
	which only rescales the partially normally ordered fluctuations
	$\langle {\stackrel{A}{:}}
			(\Delta\hat L)^2
	{\stackrel{A}{:}} \rangle=\langle (\Delta \hat L)^2 \rangle
	-\langle \hat b^\dag\hat b\rangle$,
	without affecting the sign.

\subsection{Noise}

	Analogously to the above loss scheme, excess noise is modeled with $\hat a\mapsto \sqrt{g}\hat a+\sqrt{g-1}\hat c^\dag$ with a bath in vacuum, where $g\geq 1$.
	Then, we obtain
	\begin{equation}
	\begin{aligned}
		\langle \hat a\rangle
		\mapsto
		\sqrt{g}\langle\hat a\rangle,
		\text{ }
		\langle \hat a^2\rangle
		\mapsto
		g \langle \hat a^2\rangle,
		\text{ }
		\langle \hat a\hat a^\dag\rangle
		\mapsto
		g\langle \hat a\hat a^\dag\rangle,
		\\
		\text{and }
		\langle \hat a^\dag\hat a\rangle
		\mapsto
		g\langle \hat a^\dag\hat a\rangle+g-1,
	\end{aligned}
	\end{equation}
	where the $g-1$ term describes the extra noise.
	Again, we apply this to the fluctuation of the SI field under study and find
	\begin{equation}
	\begin{aligned}
		\langle(\Delta\hat L)^2\rangle
		\mapsto{}&
		g\langle(\Delta\hat L)^2\rangle+(g-1)\langle\hat b\hat b^\dag\rangle
		\\
		\text{$\Rightarrow$ }
		\langle(\Delta\hat L)^2\rangle-\langle\hat b^\dag\hat b\rangle
		\mapsto{}&
		g\left(
			\langle (\Delta \hat L)^2 \rangle - \langle\hat b^\dag\hat b\rangle
		\right)
		\\
		{}&
		+(g-1)
		\left(
			\langle \hat b\hat b^\dag \rangle
			+ \langle \hat b^\dag\hat b \rangle
		\right),
	\end{aligned}
	\end{equation}
	scaling the
	$\langle {\stackrel{A}{:}}
			(\Delta\hat L)^2
	{\stackrel{A}{:}} \rangle$
	contribution and adding a positive term, thus not falsely decreasing the partially normally ordered fluctuations through noise.
